\documentclass[conference]{IEEEtran}
\IEEEoverridecommandlockouts
\usepackage{array}
\usepackage{cite}
\usepackage{amsmath,amssymb,amsfonts}
\usepackage{algorithmic}
\usepackage{graphicx}
\usepackage{textcomp}
\usepackage{xcolor}
\usepackage{multirow}
\usepackage{footnote}
\def\BibTeX{{\rm B\kern-.05em{\sc i\kern-.025em b}\kern-.08em
    T\kern-.1667em\lower.7ex\hbox{E}\kern-.125emX}}
\begin{document}
\setcounter{footnote}{-1}
\title{Req2Lib: A Semantic Neural Model for Software Library Recommendation}
\author{\IEEEauthorblockN{Zhensu Sun, Yan Liu\thanks{\IEEEauthorrefmark{4}Yan Liu is corresponding author.}\IEEEauthorrefmark{4}, Ziming Cheng, Chen Yang, Pengyu Che}
\IEEEauthorblockA{\textit{School of Software Engineering} \\
\thanks{This work is supported by Intelligent City Knowledge Service System (http://icity.ikcest.org/).}
\textit{Tongji University}\\
Shanghai, China \\
\{87su, yanliu.sse, 1750114, 1610833, chepengyu\}@tongji.edu.cn}
}
\maketitle

\begin{abstract}
Third-party libraries are crucial to the development of software projects. To get suitable libraries, developers need to search through millions of libraries by filtering, evaluating, and comparing. The vast number of libraries places a barrier for programmers to locate appropriate ones. To help developers, researchers have proposed automated approaches to recommend libraries based on library usage pattern. However, these prior studies can not sufficiently match user requirements and suffer from cold-start problem. In this work, we would like to make recommendations based on requirement descriptions to avoid these problems. To this end, we propose a novel neural approach called Req2Lib which recommends libraries given descriptions of the project requirement. We use a Sequence-to-Sequence model to learn the library linked-usage information and semantic information of requirement descriptions in natural language. Besides, we apply a domain-specific pre-trained word2vec model for word embedding, which is trained over textual corpus from Stack Overflow posts. In the experiment, we train and evaluate the model with data from 5,625 java projects. Our preliminary evaluation demonstrates that Req2Lib can recommend libraries accurately.
\end{abstract}


\section{Introduction}

Today, third-party libraries play an important role in modern software projects. With them, developers can improve the quality of software project and do not need to "re-implement the wheel". However, there are over 4.4m unique open-source packages on Libraries.io\cite{jeremy_katz_2018_2536573} who collects data from package managers and source code repositories. The availability of such a huge amount of reusable libraries facilitates
software development and evolution.
It also places a barrier for developers to locate useful ones. 

Usually, developers search on the Internet based on traditional keyword matching to retrieve libraries. The names of suitable library may be shown in the websites like blogs or forums whose content are not well-organized. Thus, developers have to manually examine many web pages to pick the most appropriate libraries during the process of retrieval. On the one hand, it requires a certain level of experience of related domain. On the other hand, the retrieval consumes considerable energy and time of developers\cite{kevic2014automatic}. Thus, it is difficult and time-consuming to find the appropriate ones without additional guides or experiences.

To save developers from filtering among search results, researchers have proposed models that can recommend relevant libraries for software projects using project content (e.g., libraries, source codes)\cite{thung2013}\cite{Ouni2016}\cite{Katsuragawa2018}\cite{Chen2016Detecting}. Although their models can make recommendations correctly in their working contexts, they have a set of essential defects that do harm to its practicability:

\begin{itemize}

\item \textbf{Requirement insufficiently matching:} Previous studies are based on library usage pattern\cite{thung2013}\cite{Ouni2016}\cite{Katsuragawa2018} or library similarity\cite{Chen2016Detecting} to make recommendations. All of them are content-based recommendation systems. Their results have no relation with requirements. In a supermarket, customers may pay for high-quality products recommended by sellers even if the products are not on their shopping lists. However, in the domain of software engineering, projects have clear requirements, and libraries that are irrelevant to the requirement will not be "paid". Therefore, recommendations without requirements may not help the development of software projects.

\item \textbf{An awkward situation caused by cold-start problem:} Previous study suffers from the cold-start problem: it fails when no project content are available. The content of a project need to accumulate for some time before it is enough for an effective recommendation. Thus, their model mainly works during software maintenance and evolution. However, research indicates that projects do not appear to change their dependencies over time\cite{Zerouali2018}. Because of the fear of system errors, developers are not willing to change libraries ("if it ain't broke, don't fix it"). In other words, it is not especially necessary to recommend libraries for projects that are already implemented. 
\end{itemize}

Considering the above issues, a potential way is to recommend relevant libraries directly from requirements. In the process of software development, project requirements are often identified before technology selection. In this way, developers can get recommendations while technology selection. On the one hand, technology selection is the phase that suggestions on libraries are urgently needed. On the other hand, using requirement descriptions, developers do not need to accumulate enough project content. Thus, by directly utilizing requirement information, recommendations can match the requirements of projects without cold-start problem.

To this end, we conduct an exploratory attempt to identify suitable libraries from a description of requirement. Recommendations based on natural language (NL) is beyond the capability of traditional recommendation algorithms like collaborative filtering. To exploit NL, previous researchers treat sentences as bags-of-words (BOW) and try to match the textual similarity\cite{Ouni2016}\cite{Thung2013a}. BOW lack an deep understanding of the semantics of NL. A BOW model can not tell the difference between "before long" and "long before". 

In this work, we propose an approach based on deep learning, called Req2Lib, which can recommend libraries given a NL description of the requirements. Req2Lib adapt a neural model called Sequence-to-Sequence (seq2seq)\cite{sutskever2014sequence}. It is trained with a dataset of historical repository content, which contains two parts: requirement descriptions and libraries listed in the configuration files. In addition, we apply a domain-specific embedding model\cite{Efstathiou2018} trained over a corpus of Stack Overflow posts to represent words with high-dimensional vectors. With domain-specific embedding, Req2Lib can get more domain knowledge from the embedded input. 
 
In the experiment, we train and evaluate Req2Lib using the descriptions and libraries of 5,625 unique projects. Finally, the experiment results show that Req2Lib can recommend libraries accurately with recall rate@10 of 0.904, precision@10 of 0.316, respectively. 

The contributions of our work are as follows:
\begin{itemize}
  \item We identify a new perspective on the domain of library recommendation which helps to solve the defects above: Recommend useful libraries given requirement descriptions of a project.
  \item We propose a deep learning approach to accurately recommend available third-party libraries from requirements, which proves the practicality of our perspective.
\end{itemize}

\begin{figure*}[htbp]
\centerline{\includegraphics[width=16cm]{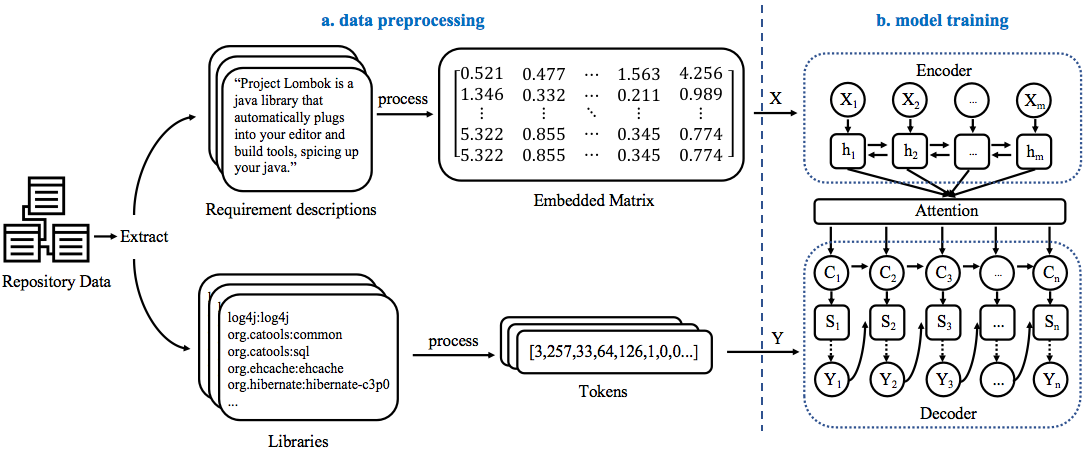}}
\caption{Overview of Req2Lib. The dotted lines denote the masked softmax layer}
\label{overview}
\end{figure*}

\section{Related Work}
Among the previous studies of the recommendation system at the library-level, it is common to use the content of projects (i.e. current library usage, source code) to make recommendations. LibRec\cite{thung2013}, proposed by Thung \textsl{et al.}, use association rule mining and collaborative filtering to recommend libraries. LibFinder\cite{Ouni2016}, proposed by Ouni \textsl{et al.}, improves the performance of LibRec by recommending libraries based on linked-usage of libraries and semantic similarity of identifiers of source code. DSCRec\cite{Katsuragawa2018} introduces domain-specific categories to the LibRec. Instead of using historical data of projects, Chen \textsl{et al.}\cite{Chen2016Detecting} constructed an analogical-library knowledge base by mining Q\&A data from Stack Overflow to calculate the similarity between libraries. However, studies above can not provide solutions to the their defects (i.e., requirement insufficiently matching and cold-start problem).

Our approach differs from all the previous works in the fact that we take the requirement descriptions instead of project libraries as input. To the best of our knowledge, this paper is the first to investigate the use of requirement descriptions to address the problems of existing studies.

\section{Req2Lib}
\subsection{Overview}
Req2Lib takes natural language as input and generates a sequence of  related libraries. When the model recommends a library, it considers not only the semantic information from input but also the correlation information from the libraries recommended before. In addition, it contains a deep understanding of natural language combined with domain knowledge from the embedding model. Fig.~\ref{overview} shows the overall framework of our approach. It comprises two steps: a) data preprocessing; b) model training.

\subsection{Preprocessing}

\subsubsection{Data Collection}
In this step, we collect project requirement descriptions and the corresponding libraries. Particularly, our model utilizes semantic information in NL so that we can use various forms of texts in natural language as long as it contains requirement information (e.g., requirement documents, project description, readme document). On the one hand, the proprietary of requirement documents limits its acquisition. On the other hand, readme documents are usually not well-organized which contains many unrelated information. Thus, we choose the project descriptions as the substitute of requirement documents. We get the data required by Req2Lib from Libraries.io.

\subsubsection{Data Processing}
\begin{figure}[htbp]
\centerline{\includegraphics[width=6.5cm]{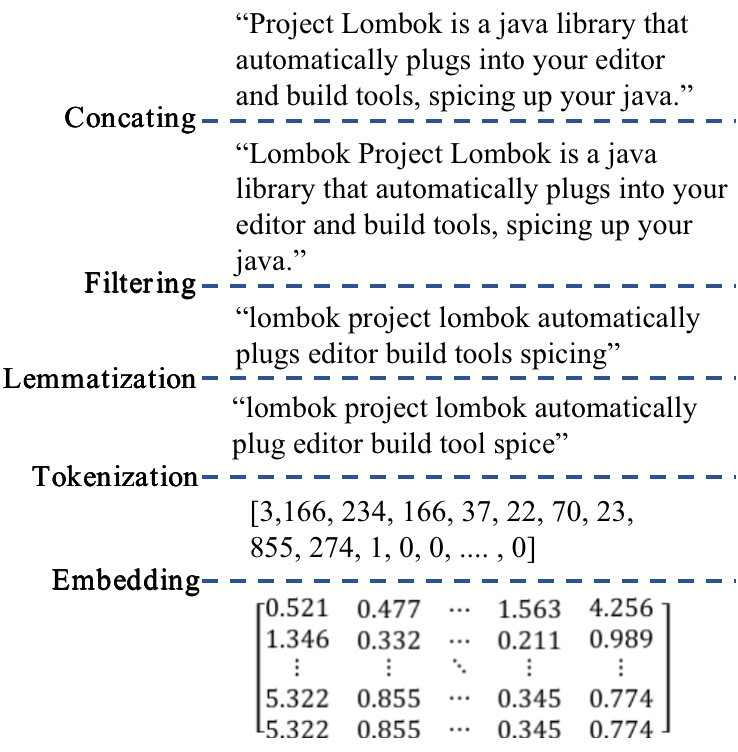}}
\caption{An example of description processing.}
\label{example1}
\end{figure}

The data we collected have poor quality. Thus, we apply several processing methods to the dataset. An example of the procedure for processing NL is shown in Fig.~\ref{example1}.

\textbf{Concatenating:} The name of a project contains semantic information. To get more information, we splice the project names and stitch them before the input texts.
   
\textbf{Filtering:} For natural language, we turn all the words into lowercase and remove the stop words\footnote{https://www.textfixer.com/tutorials/common-english-words.txt} and special symbols. At the same time, we use a vocabulary\cite{Efstathiou2018} based on Stack Overflow Q\&A posts to filter out words that are not related to the software domain. For libraries, we remove low-usage libraries because libraries adopted by few people may have quality defects.
   
\textbf{Lemmatization:} Using the Stanford Core NLP tool\footnote{https://stanfordnlp.github.io/CoreNLP/}, we reduce the words to its base form. For instance, the word 'libraries' will be replaced by the word 'library'. In this way, a word in different forms can share the same weights in our model. 

\textbf{Sorting:} The deep learning model that we use is a sequential model. To improve the performance, libraries should be listed in a fixed order. Thus, we sort the libraries of each project descending by library frequency in the total data.

\textbf{Tokenization:} We tokenize the desriptions and obtain a sequence of tokens. As for libraries, each library is treated as a token. With randomly generated vocabularies, we represent the tokens with numerical IDs. In this way, we turn the natural languages and libraries into sequences of numerical tokens with fixed length. 
   
\textbf{Embedding:} Using neural models, we can present the words by real-valued vectors. Our model utilizes a domain-specific pre-trained word2vec model\cite{Efstathiou2018} that is trained over textual corpus from Stack Overflow posts. The word2vec model\cite{mikolov2013efficient} can captures detailed relational structure, and a broad range of topics in the field. With the pre-trained model, we convert the tokens into 200-length high-dimensional vectors that contains related software domain knowledge.

After the above processing steps, we convert the descriptions and libraries into a sequence of dense embedding vectors and a sequence of tokens separately.

\subsection{Sequence-to-Sequence}
The library recommendation task can be treated as a multi-label classification problem by taking libraries as labels. Thus, we implement our model based on the work of Yang et al\cite{Yang2018}. They proposed a variant of the seq2seq model with a novelty decoder for multi-label classification. The overview structure is shown in the model training part of Fig.~\ref{overview}.

\textbf{Encoder:} We adopt a deep neural network called Long Short-Term Memory (LSTM)\cite{hochreiter1997long}. To fully utilize the context of input, we use a bi-direction LSTM as the encoder to extract semantic information from the input in both forward and reverse directions. At each step $t$, it reads the $t$-th embedded token $x_t$ from the input sequence, then computes the hidden states $h_t$, namely,
\begin{equation}
\overrightarrow{h_t} = \overrightarrow{LSTM}(x_t, \overrightarrow{h_{t-1}});\overleftarrow{h_t} = \overleftarrow{LSTM}(x_t, \overleftarrow{h_{t+1}}) 
\end{equation}

By concatenating $\overrightarrow{h_t}$ and $\overleftarrow{h_t}$, we get the final output of the Encoder, a sequence of hidden state $h_1, h_2,\cdots, h_n$ called context vector. It represents the semantic information and other latent features extracted from the embedded input.

\textbf{Attention:} For the contributions of each keyword to the recommended result is different, we use an attention component\cite{bahdanau2014neural} to score each hidden state of encoder and compute the weighted average to generate new context vectors. The weights ${\alpha}_{ti}$ to the $i$-th word of time step $t$ is computed as follow:
\begin{equation}
e_ti = v_a^Ttanh(W_as_t+U_ah_i);{\alpha}_{ti} = \frac{exp(e_{ti})}{\sum_{j=1}^mexp(e_{tj})}
\end{equation}
where ,$W_a$,$U_a$,$v_a$ are weights parameters of linear layers and $s_t$ is the current hidden state of decoder. With weights, the new context vectors are computed as follow:
\begin{equation}
c_t = \sum_{i=1}^m\alpha_{ti}h_i
\end{equation}

\textbf{Decoder:} To predict the current results from the context vector $c_t$, we use a LSTM model as the decoder. It takes the context vector, previous prediction, and previous hidden state as input. In other words, it take use of semantic information and library linked-usage information. Its hidden state $s_t$ is computed by:
\begin{equation}
s_t = LSTM([emb(y_{t-1});c_{t-1}],s_{t-1})
\end{equation}
where $[emb(y_{t-1});c_{t-1}]$ is the concatenation of $emb(y_{t-1})$ and $c_{t-1}$. $emb(y_{t-1})$ is the embedding of a library whose id has the biggest value in $y_{t-1}$, $y_{t-1}$ is the probability distribution of the model output in the previous step, which is computed as follow:
\begin{equation}
o_t = W_orelu(W_ds_t+V_dc_t);y_t = softmax(o_t+I_t)
\end{equation}
where $W_o$,$W_d$,$V_d$ are the weight parameters of the linear layers. $I_t$ is the mask at time step $t$ to prevent the decoder predicting a repeated library $i$, which is calculated by: 
\begin{equation}
(I_t)_i = \begin{cases}
- \infty &\mbox{if $i$ has repeated at previous steps}\\ 
0 &\mbox{otherwise}\\
\end{cases}
\end{equation}

\textbf{Loss:} The goal of the model is to minimize the value of loss function. On the basis of cross-entropy loss, we apply a popularity-based coefficient to the weight of each item. Thus, we use the weighted cross-entropy loss as the loss function of our model. The coefficients are computed by:
\begin{equation}
L(y) = - \frac{1}{N}\sum_{i=1}^N\sum_{j=1}^nw_jlogp(y_j^i); w_j = 1 - \frac{f_j}{\sum_{k=1}^nf_k}
\end{equation}
where $N$ is the total number of a training batch, $n$ is the number of target libraries of each project, $f_j$ is the frequency of a library whose ID is $j$.

Cross-entropy loss is widely used in classification models. The reason why we use weighted one is to set barriers to popular libraries. Among the libraries used by developers, there are many popular ones that are acceptable for almost all projects (e.g., Junit). The model can always recommend these common libraries to reduce the loss value. However, from the user's point of view, predictable recommendations are not particularly useful. The predictable recommendation is a new problem, to which previous studies do not provide any solutions.  

\section{Experiments}
\subsection{Dataset}
Our study uses the dataset extracted from Libraries.io\cite{jeremy_katz_2018_2536573} on 22 Dec 2018. To improve the data quality, we filter the project with the following criteria:
\begin{itemize}
  \item The project should use Java as the programming language with over ten stars. Projects with low stars may have problems in quality.
  \item Following the work of Thung et al\cite{thung2013}, the project should use at least ten libraries. 
  \item To filter out the descriptions that contains little valid information, the project should have a piece of descriptive text which contains more than three words. 
  \item The project should be unique.
\end{itemize}
Finally, we obtained 5,625 pieces of data. The collected projects contain 1468 libraries and . 80\% of them are randomly picked as training sets, and the rest are used as test sets.

\subsection{Evaluation Metrics}
\textbf{Recall rate@k:} Recall rate@k is widely used by previous studies to evaluate their models\cite{thung2013}\cite{Thung2013a}. It is calculated as follows:
\begin{equation}
Recall rate@k = \frac{\sum_{i=1}^{N}isFound(S_i)}{N}
\end{equation}
and
\begin{equation}
isFound(x) = \begin{cases}
1 &\mbox{$x$ is in ground truth}\\
0 &\mbox{otherwise}\\
\end{cases}
\end{equation}
where $S_i$ is the $i$\_th model result.

\textbf{Precision@k:} Precision@k is defined as the proportion of ground truth hits among top-k model results. It is calculated by:
\begin{equation}
Precision@k = \frac{TP}{TP + FP}
\end{equation}
where $TP$ is the number of correct ones among top k model results, $FP$ is the number of incorrect ones among top k model results.

\textbf{Popularity-Stratified Recall@k} Metrics of previous studies mainly focus on the accuracy of models. To assess predictable recommendations, we adopt popularity-stratified recall (PSR)@k\cite{Steck2011}. On the basis of classical recall, PSR introduces the inverse response probability as the weights of each library. In this way, popular libraries acquire a lower weight. It is calculated by:
\begin{equation}
PSR@k = \frac{\sum_{i\in S_u^{+,k}}s_i}{\sum_{i\in S_u^{+}}s_i};s_i \propto 1/(N_{obs,i}^+)^\beta \quad 
\end{equation}
where $S_u^{+,k}$ is the sequence of recommended results for project $i$, $S_u^{+}$ is the sequence of ground truth for project $i$, $N_{obs,i}^+$ is the frequency of a library $i$ among all the projects. We set $\beta$ to 0.2 to prevent the models who only recommend unpopular items from winning.
\subsection{Details}
According to the above methods, we obtain the dataset and train the model. During training, we use the Adam optimization algorithm\cite{kingma2014adam} to minimize the weighted cross-entropy loss. To improve the accuracy, we adopt several tricks on seq2seq:

\textbf{Dropout regularization\cite{srivastava2014dropout}:} The key idea of dropout to randomly drop units from the neural network during training. It can prevent neural network from over-fitting.

\textbf{Gradient clipping\cite{pascanu2013difficulty}:} This strategy limit the gradient to a maximum number, which helps to deal with exploding gradients and vanishing gradients problem

\textbf{Beam search\cite{wiseman2016sequence}:} It is applied for measuring sub-optimal recommendation. In this way, we can get the global optimal result.

\subsection{Results}
\begin{figure}[htbp]
\centerline{\includegraphics[width=6cm]{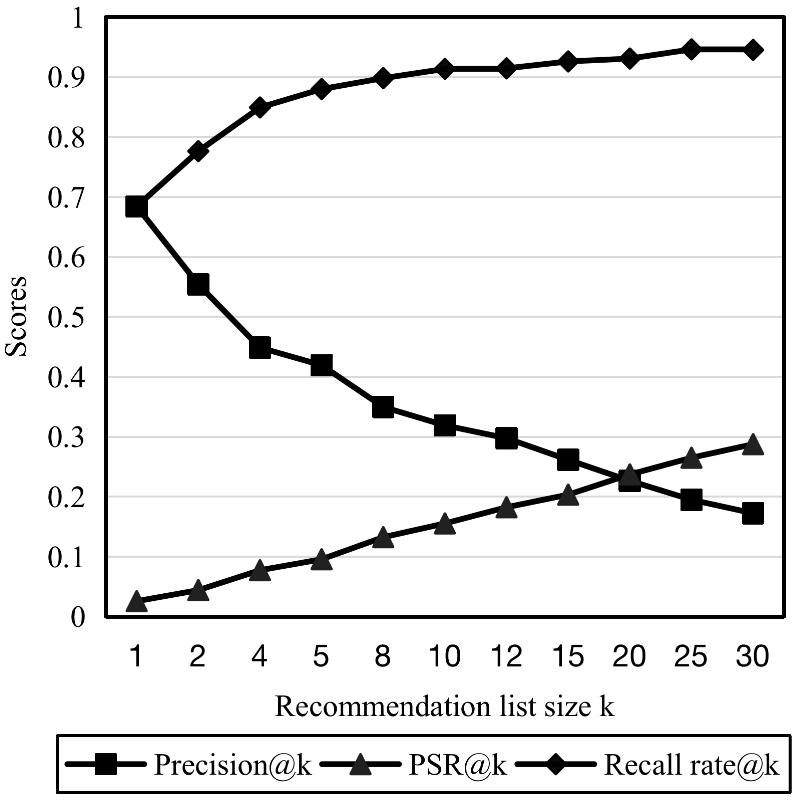}}
\caption{Different performance on varying recommendation list sizes.}
\label{effect}
\end{figure}

With the experimental results obtained from the model on the test set, we compare the recommended libraries with actually used libraries and calculate the evaluation metrics with different recommendation size $k$. The results is shown in Table.~\ref{result_table} and Fig.~\ref{effect}. 

For the recall rate@k, with the increase of recommendation list sizes $k$, the performance of our method for recall rate@k has improved. When the size reaches 10, our model can recommend at least one useful library for more than 90\% of projects. For the precision@k, a more significant k results in a lower precision, because confident results have a higher priority in our model. Req2Lib do not perform well on PSR@k for the reason that PSR is a fairly strict criterion.

\begin{table}[htbp]
\caption{Obtained evaluation results}
\renewcommand{\arraystretch}{1.3}
\begin{center}
\begin{tabular}{ccccccccccccccc}
\hline
\textbf{Metrics} & k=1 & k=5 & k=10 & k=20\\
\hline
Recall rate@k & 0.684 & 0.880 & 0.904 & 0.931\\
Precision@k & 0.684 & 0.419 & 0.316 & 0.227\\
PSR@k & 0.026 & 0.096 & 0.156 & 0.237\\
\hline
\end{tabular}
\label{result_table}
\end{center}
\end{table}

\subsection{Threats to Validity}
The results can not present the price for changing the way of making recommendation. The reason is that there are no similar studies on requirement-based library recommendation. Furthermore, it is hard for a sequential model to compare with previous studies\cite{thung2013}\cite{Ouni2016} directly. In their experiments, for a project, half of its libraries are used as model input. The rest need to be kept for evaluation. Different from them, our model recommend all libraries given requirement descriptions. As a sequential model, the first recommended library will influence the choice of next library. Thus, evaluating half of them can not present the actual accuracy of our approach. A deeper study is needed.

\section{Conclusion}
This paper presents our new perspective on library recommendation to avoid the essential defects of previous studies. We also propose a deep learning approach called Req2Lib to recommend libraries given requirement descriptions. The experimental results of Req2Lib demonstrate that the perspective we identified is practical. Besides, the promising results also indicate that it worth a further study for requirement-based library recommendation using deep learning. 


\end{document}